# Efficient spectrum prediction and inverse design for plasmonic waveguide systems based on artificial neural networks

Tian Zhang,[1] Jia Wang,[1] Qi Liu,[1] Jinzan Zhou,[1] Jian Dai,[1] Xu Han,[2] Yue Zhou,[1] and Kun Xu[1,*]

[1]State Key Laboratory of Information Photonics and Optical Communications, Beijing University of Posts and Telecommunications, Beijing 100876, China
[2]Huawei Technologies Co., Ltd, Shenzhen 518129, Guangdong, China
*Corresponding author: xukun@bupt.edu.cn

In this article, we propose a novel approach to achieve spectrum prediction, parameter fitting, inverse design and performance optimization for the plasmonic waveguide coupled with cavities structure (PWCCS) based on artificial neural networks (ANNs). The Fano resonance and plasmon induced transparency effect originated from the PWCCS have been selected as illustrations to verify the effectiveness of ANNs. We use the genetic algorithm to design the network architecture and select the hyper-parameters for ANNs. Once ANNs are trained by using a small sampling of the data generated by Monte Carlo method, the transmission spectrums predicted by the ANNs are quite approximate to the simulated results. The physical mechanisms behind the phenomena are discussed theoretically, and the uncertain parameters in the theoretical models are fitted by utilizing the trained ANNs. More importantly, our results demonstrate that this model-driven method not only realizes the inverse design of the PWCCS with high precision but also optimizes some critical performance metrics for transmission spectrum. Compared with previous works, we construct a novel model-driven analysis method for the PWCCS which are expected to have significant applications in the device design, performance optimization, variability analysis, defect detection, theoretical modeling, optical interconnects and so on. © 2018 Chinese Laser Press

https://doi.org/XX.XXXX/PRJ.X.XXXX

## 1. INTRODUCTION

Owning to the unique properties of near field enhancement effect and breaking the diffraction limit, the emergence of surface plasmon polaritons (SPPs) has attracted a great deal of research attentions [1]. Until now, diversified plasmonic structures have been proposed to excite and transmit the SPPs, such as metamaterial [2, 3], dielectric grating and metallic grating [4, 5], metal-dielectric-metal (MDM) waveguide [6-9], graphene-based waveguide [10, 11], hybrid waveguide [12-14] and so on. In these structures, the plasmonic waveguide coupled with cavities structures (PWCCSs) which can be easily integrated into the plasmonic circuits has attracted widespread attention because it is at subwavelength scale, supports relatively long propagation length for SPPs and demands relatively simple fabrication by using electron beam lithography and focused ion beam etching [14-16]. As for the simple PWCCSs, the physical mechanisms behind the phenomena are analyzed by utilizing some theoretical models and classical methods such as coupled-mode theory (CMT), transfer matrix method (TMM) and so on [6-9, 17]. Then the theoretical models are constructed to predict transmission spectrum, determine structure parameters and optimize some critical metrics (transmittance and bandwidth) [6-9]. However, for the relatively complex PWCCSs with complicated waveguide and cavity structure, the physical mechanism is hardly understood and thus theoretical models are difficult to construct [18, 19]. And the absence of empirical relationship between the structure parameters and electromagnetic responses often enforces utilization of the time-consuming brute force search or evolutionary algorithms to determine the shape, dimension and variability of the device [20]. Obviously, an effective intelligence algorithm which obtains reliable spectrum prediction, inverse design and performance optimization should be addressed in the design and analysis of photonic devices.

For the complex PWCCSs, computing the electromagnetic responses for all structure parameters via numerical simulation methods usually requires tremendous computation time. If the electromagnetic responses for all structure parameters can be predicted by using a small sampling of simulation results, the efficiency of design and analysis for the complex PWCCSs will be improved. However, a simple and quick solution to predict and evaluate the spectrum responses for all structure parameters based on the partial simulation results is still lacking. In addition, although inverse design and performance optimization have been used to assist the design of mode multiplexer [21], wavelength multiplexer [22], polarization beam splitter [23], polarization rotator [24], power splitter [25] and so on, few studies have been focused on the PWCCSs. Generally speaking, inverse design and performance optimization problems are solved by using several optimization algorithms, such as evolution algorithm (genetic [24] and particle swarm [25]), topology optimization [23], adjoint method [22], convex optimization [22], nonlinear search method [23] and so on. Among these optimization algorithms, the genetic algorithm (GA) is widely used because of effectiveness, simplicity and intuitiveness, even it requires a lot of time to evolve, crossover and mutate [26]. And even these optimization algorithms can iteratively optimize for some specific metrics, they are difficult to directly achieve the most suitable structure parameters for a complete transmission spectrum in a wide wavelength range. In recent years, artificial neural networks (ANNs) has been applied in approximating many physics phenomenon with high degrees of precision [27-34]. For example, the quantum many-body problem could be solved by utilizing ANNs [27]. Y. Shen et al. pointed out that the trained ANNs could be used to simulate the light scattering of multilayer nanoparticles with different thicknesses [28]. And the trained ANNs could solve the spectrum prediction and inverse design problems more quickly than numerical simulation method [28, 29]. In order to avoid the data inconsistency problem in the inverse design for photonic devices, a tandem network structure which composed of forward-modeling unit and inverse-design unit was proposed [30]. And ANNs-based numerical methods has been proposed to design and optimize the complex photonics devices, for example power splitter [31], meta-gratings [20] and meta-materials [32, 33]. Other machine learning algorithms, such as reinforcement learning and perceptron algorithm, were used to design a subwavelength optical coupling device [34]. It should be noted that the design of neural network architectures and the selection of hyper-parameters for ANNs requires a lot of expert knowledge [35]. Lately, GA [36], Bayesian optimization [37, 38] and reinforcement learning [38] were tried for the automated design of ANNs. However, few studies in the above-mentioned works introduce the design process of the network architectures for ANNs, which is critical for the prediction accuracy and algorithmic convergence.

In this article, we propose a novel method using ANNs to achieve spectrum prediction, inverse design and performance optimization for the PWCCSs. To verify the effectiveness of ANNs, the Fano resonance (FR), especially for the plasmon induced transparency (PIT) effect, originated from the mode coupling in the PWCCSs are taken into consideration. We use the GA to design the network architectures and select the suitable hyper-parameters for ANNs. It is important to note that the transmission spectrum predicted by ANNs are approximate to the finite-difference time-domain (FDTD) simulated results with high precision. In addition, the physical mechanisms behind the FR and PIT effects are discussed based on the CMT and TMM, and the uncertain parameters in the theoretical models are fitted by using the trained ANNs effectively. Moreover, the ANNs have been successfully employed in solving the inverse design and performance optimization problems for the PWCCSs.

## 2. DEVICE DESIGN AND SIMULATION RESULTS

It has been demonstrated that the FR and PIT effect can be found in the transmission spectrum of the PWCCSs due to the mode coupling between the wideband bright modes and narrowband dark modes [6-9]. The PIT effect is often regarded as a special case of the FR whose spectrum line shape around the transmission peak is asymmetric [2]. Two different coupling methods are used to explain the FR and PIT effect in the PWCCSs: one is based on the direct near-field coupling between bright modes and dark modes [8, 39, 40], the other is based on the indirect destructive interference through waveguide shift coupling [6, 7, 9]. Correspondingly, the physical mechanisms of the FR and PIT effect can be explained by the destructive interference between two pathways in a three-level atomic system including the ground, excited and metastable states, or, equivalently, the doublet of dressed states [39]. In this article, we construct three different PWCCSs which include different number of cavities as illustrations to verify the effectiveness of ANNs. Fig. 1(a) exhibits the simplest three-resonators-coupled (THRC) system which consists of a MDM waveguide and three side-coupled comb cavities. Compared with the THRC system [Fig. 1(a)], another one and two rectangular cavities are added in the up side of the cavities 1, 2 and 3 to construct a four-resonators-coupled (FORC) system [Fig. 1(b)] and a five-resonators-coupled (FIRC) system [Fig. 1(c)], respectively. The detailed structure parameters of all PWCCSs and the detailed simulation settings of the FDTD method are described in Appendix A.

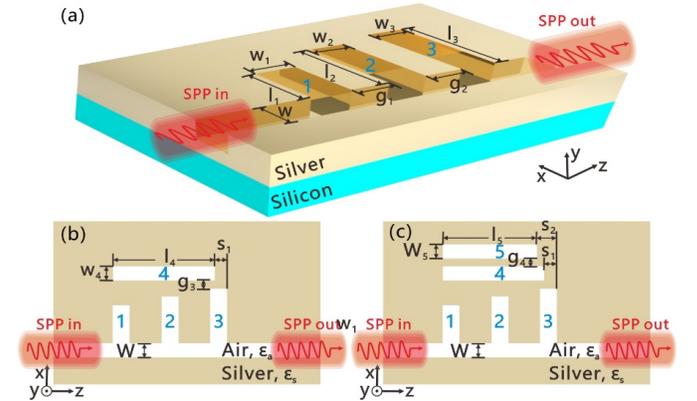

**Fig. 1.** Schematic diagrams of the (a) THRC system, (b) FORC system and (c) FIRC system.

When TM polarized SPPs are injected from the left port of the THRC system, the propagating plasmonic waves confined to the metal-dielectric interface can directly couple into the three comb cavities [7]. As shown in Fig. 2(a), we can observe that two obvious transmission peaks which are indicated by B and D points exist in the transmission spectrum. It is noteworthy that dips are located on the both sides of the peaks distinctly, which indicates the double PIT effects emerge in the transmission spectrum [7]. In order to get insight into the physics mechanism of the double PIT effects, the normalized magnetic field distributions of the transmission peaks and dips indicated by B, D and A, C, E are exhibited in Fig. 2. It can be found that it's the waveguide phase coupling between the cavities that gives rise to the peaks in the double PIT effects, while the reason for the appearance of the dips is related to the resonance of the cavities [6-9]. The theoretical results shown in Fig. 2(a) are calculated by using the Eq. (B11) in Appendix B based on the CMT and TMM. It can be seen that the theoretical results agree with that simulated from FDTD method basically. Notably, the suitable parameters ($\omega_1$=352.9 THz, $\omega_2$=314.1 THz, $\omega_3$=288.7 THz, $\gamma_1$=38 THz, $\gamma_2$=109 THz, $\gamma_3$=80 THz) in Eq. (B11) are fitted by using the ANNs, and the detailed principle is presented in the next section. In addition, due to the extreme dispersion in the FR and PIT effect, the slow light which is characterisced by the group index $n_g=(c\times\tau_g)/D$ $=(c/D)\times(d\psi(\omega)/d\omega)$ is shown in Fig. 2(b) [6-9]. Here, $c$ is the light velocity in vacuum, $\tau_g$ is the group delay, $D$=1100 nm is the length between the source and monitor, and $\psi(\omega)$ is the transmission phase shift [9]. It can be observed that two maximum group indices 6.04 and 7.74 are achieved for the double PIT effects at the transparency peak wavelengths 874 nm and 984 nm, respectively. Furthermore, we also calculate the dephasing times for the double PIT effects via $T_r= 2\hbar/\Gamma$, where $\hbar$ is the reduced Planck's constant and $\Gamma$ is the full width at half maxima (FWHM) of the PIT effects [41, 42]. For the THRC system, the dephasing times of the transmission peaks on the left (B) and that on the right (D) are estimated as 0.35 ps and 0.45 ps, respectively.

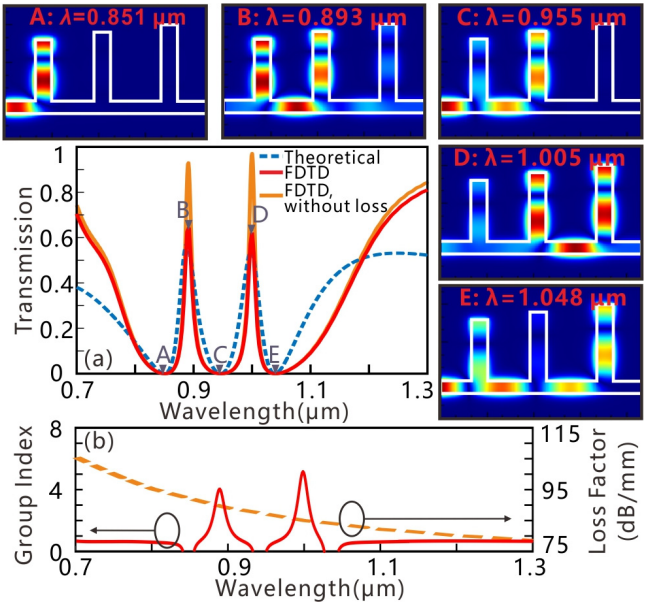

**Fig. 2.** (a) Simulated transmission spectrum of the THRC system for silver with loss (red solid line) and without loss (orange solid line), and theoretical transmission spectrum of the THRC system (blue dashed line). (b) Group index and loss factor of the THRC system. The insets are simulated magnetic field distributions for the incident light at wavelengths of 851 nm (A), 893 nm (B), 955 nm (C), 1005 nm (D), and 1048 nm (E).

The physical mechanism of the double PIT effects in the THRC system is relatively simple, which only takes the waveguide phase coupling into consideration. By contrast, we propose two relatively complex PWCCSs which include the direct near-field coupling and indirect waveguide coupling simultaneously. In the FORC [Fig. 1(b)] and FIRC [Fig. 1(c)] systems, the rectangular cavities newly added in the structures are regarded as dark modes because they are excited by the comb cavities (bright mode) rather than the bus waveguide [40]. Here, the FDTD simulated transmission spectrums (red solid line) and theoretical transmission spectrums (blue circles) for the FORC and FIRC systems are depicted in Fig. 3(a) and Fig. 4(a), respectively. Compared with the FDTD simulated results in Fig. 2(a), the optical characteristics around 1.18 μm in Fig. 3(a) and Fig. 4(a) become steep and asymmetric, indicating the appearance of the FRs [43]. Interestingly, the double PIT effects and the FRs simultaneously appear in the transmission spectrum, which is rarely mentioned in the related articles [6-9]. For the FRs in Fig. 3(a) and Fig. 4(a), the phase is dramatically changed [the transmittance varies sharply from the peak to dip with a small wavelength range of 12 nm (FORC) and 6 nm (FIRC)], which is suitable for the application of switches, sensors, slow light and so on [44]. As shown in Fig. 3(b) and Fig. 4(b), the maximum group indices for the FORC and FIRC systems are 9.84 and 7.21, respectively. In addition, the dephasing times of the PIT peaks in the FORC and FIRC systems are similar to those in the THRC system because of the similar FWHM (15~20 nm). Compared with the dephasing time of the single FR dip in the FORC system (0.42 ps), the double FR dips in the FIRC system have relatively larger values $T_G$=0.95 ps and $T_I$=0.61 ps due to the smaller FWHM. Obviously, the calculated dephasing times in this article are larger than the general dephasing times of FR (on the order of 10 fs) [41, 42].

In order to analyze the physical mechanism of the FR and PIT effects in the FORC system, the corresponding magnetic field distributions are shown in Fig. 3, where the plasmonic modes in the rectangular cavity are excited for the peak F and dip G collectively. In Fig. 3(a), the transmission spectrums for the PWCCSs which include only cavities 1, 2, 4 (orange dash line) and cavities 1, 2 (blue dash line) are identical. In addition, the FR becomes weak when coupling distance $g_3$ increases from 20 nm to 30 nm, while other peaks and dips are stable. We can infer that it's the destructive interference between the rectangular cavity 4 and comb cavity 3 gives rise to the transmission peak F because the near-field coupling between the cavities 1, 2 and 4 is negligible. More importantly, the theoretical transmission spectrum calculated by using Eq. (B15) in Appendix B is quite approximate to the FDTD simulated results. The fitted parameters in Eq. (B15) are $\omega_1$=352.9 THz, $\omega_2$=314.1 THz, $\omega_3$= 288.7 THz, $\omega_4$=255.7 THz, $\gamma_1$= 38 THz, $\gamma_2$=109 THz, $\gamma_3$=80 THz and $\gamma_4$=0.08 THz. For the FIRC system, the physical mechanism of the double FRs in Fig. 4(a) is similar to the single FR shown in Fig. 3(a), whereas the difference is the occurrences of the dips G and I are the resonance in the cavity 5 and 4, respectively. From the magnetic field distributions F, G, H and I shown in Fig. 4(a), it can be observed that it's the destructive interference between all the rectangular cavities in the FIRC system and the comb cavity 3 forms the transmission peaks F and H, which is demonstrated by the fact that optical characteristics of the FRs become less steep when coupling distance $g_4$ is increased from 40 nm to 60 nm. Here, the theoretical results (blue dash line) shown in Fig. 4(a) are calculated by using Eq. (B21) in Appendix B. In Eq. (B21), the fitted parameters predicted by ANNs are $\omega_1$=352.9 THz, $\omega_2$=314.1 THz, $\omega_3$= 288.7 THz, $\omega_4$=255.7 THz, $\omega_5$=257.5 THz, $\gamma_1$=38 THz, $\gamma_2$=109 THz, $\gamma_3$=80 THz, $\gamma_4$=0.08 THz and $\gamma_5$=0.2 THz. Here, since we don't take the higher order and lower order resonance modes in the cavities into consideration, the theoretically calculated results imperfectly match with the FDTD simulated results.

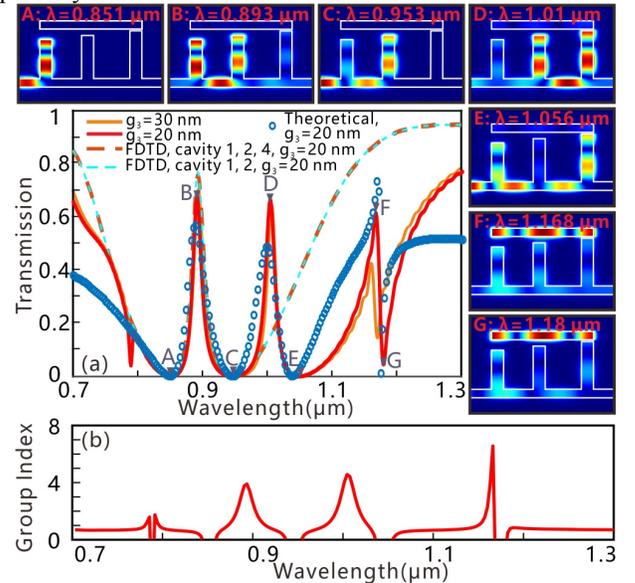

**Fig. 3.** (a) Simulated transmission spectrum of the FORC system for $g_3$=20 nm (red solid line) and 30 nm (orange solid line). Theoretical transmission spectrum of the FORC system (blue circles). Simulated transmission spectrum of the FORC system which include only cavities 1, 2, 4 (orange dash line) and cavities 1, 2 (blue dash line). (b) Group index of the FORC system. The insets are calculated magnetic field distributions for the incident light at wavelengths of (A) 0.851 μm, (B) 0.893 μm, (C) 0.953 μm, (D) 1.01 μm, (E) 1.056 μm, (F) 1.168 μm, and (G) 1.18 μm.

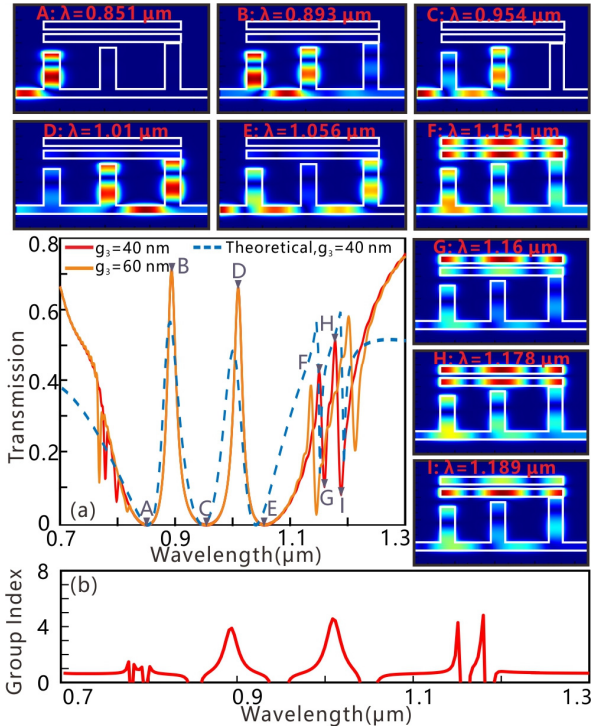

**Fig. 4.** (a) Simulated transmission spectrum of the FIRC system for $g_3$=40 nm (red solid line) and 60 nm (orange solid line). Theoretical transmission spectrum of the FIRC system (blue dash line). (b) Group index of the FIRC system. The insets are calculated magnetic field distributions for the incident light at wavelengths of 851 nm (A), 893 nm (B), 954 nm (C), 1010 nm (D), 1056 nm (E), 1151 nm (F), 1160 nm (G), 1178 nm (H) and 1189 nm (I).

## 3. SPECTRUM PREDICTION, INVERSE DESIGN AND OPTIMIZATION FOR THE PWCCS

Mining the internal relationship between all structure parameters and electromagnetic response requires high computational cost to traverse all structure parameters (brute force) or to utilize MC method [20]. The efficiency of the device design and variability analysis will be improved if all simulation results are predicable based on a small sampling of simulation results. Machine learning techniques, especially for ANNs, are data-driven methods which can predict the response for unknown data instance based on classification, clustering and regression [45]. More interestingly, it has been demonstrated that the trained ANNs can predict the same electromagnetic responses faster than conventional simulation methods [28, 29]. Here, we use ANNs to predict the transmission spectrum for arbitrary structure parameters of the PWCCSs. As shown in Fig. 5(a), the ANNs take the structure parameters (the dimension of the waveguide and cavities) as the input and predict the corresponding electromagnetic responses. For example, for the THRC system, the potential relationships between the structure parameters (the lengths, widths of the comb cavities 1, 2, 3 and the lengths of the gaps 1, 2 between the cavities) and the transmission spectrum are taken into consideration. Since the FORC and FIRC systems have more cavities than the THCR system, more structure parameters are inputted into the ANNs. The variation ranges of the structure parameters are fixed to be ±20 nm. Specifically, it means that the smallest length of the resonator 1 is 460 nm, and the largest one is 500 nm. In the FDTD simulations, the length of the resonator 1 is randomly generated from 460 nm to 500 nm with the precision of 1 nm. Repeated 2D FDTD simulations are employed to generate different 20,000 instances for 8 parameters ($l_1$, $l_2$, $l_3$, $w_1$, $w_2$, $w_3$, $g_1$, $g_2$) based on MC sampling [46]. It is noteworthy that the generation of the training and test instances including structure parameters and the discrete data points in simulated transmission spectrum requires a significant amount of time. However, the prediction process for new instance is faster than conventional simulation methods because the weights and thresholds of ANNs are fixed once the training process is completed [28]. It takes us 30 hours to generate 20000 training instances with NVIDIA Tesla P100 GPU accelerators [47]. In order to guarantee the generalization of the training models, the ANNs are trained by using the 20,000 instances, while another 2000 instances are left as the test sets to validate the training effect. The model training of ANNs is done by optimizing the mean squared error based on the stochastic gradient descent (SGD) or adaptive moment estimation (Adam). Attempting to exhibit the performance of the trained ANNs, a simple indicator, score [48]

$$1 - J^2 = 1 - \frac{\sum_{i=0}^{N}(y_{true_i} - y_{pred_i})^2}{\sum_{i=0}^{N}(y_{true_i} - y_{pred_i}/N)^2} \quad (1)$$

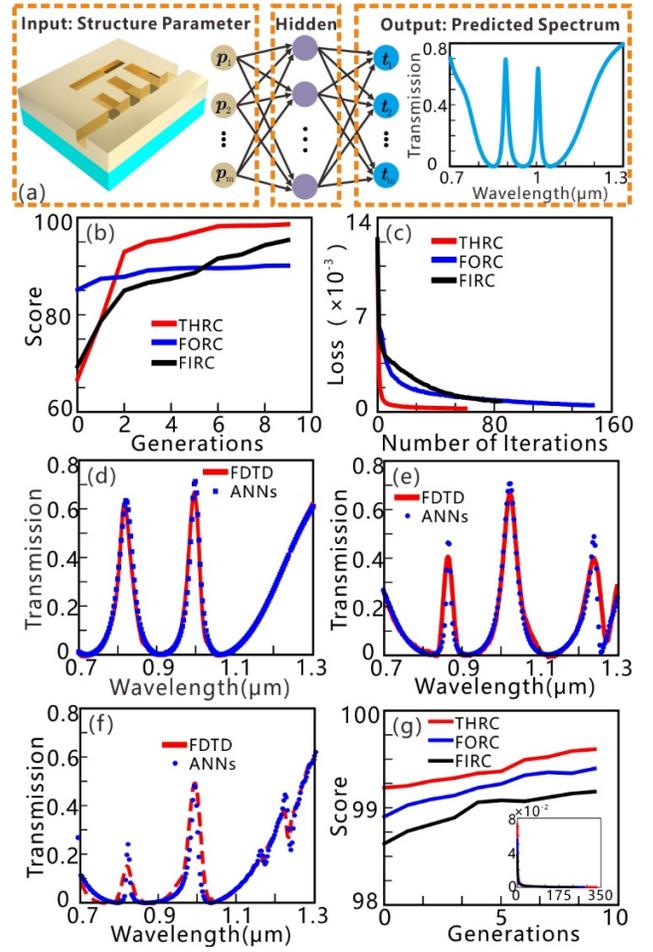

**Fig. 5.** (a) The diagram of the ANNs applied in the spectrum prediction. (b) Fitnesses for different generations in the spectrum prediction. (c) Training losses for different iterations in the spectrum prediction. The FDTD simulated transmission spectrums and ANNs-predicted transmission spectrums for the THRC (d), FORC (e) and FIRC (f) systems. (g) Fitnesses for different generations in the parameter fitting. The inset reveals the training losses for different iterations in the parameter fitting.

, is defined to measure the distance between the ANNs-predicted results and the ground truth (FDTD simulations). In Eq. (1), $N$

relates to the total discrete data points in the FDTD simulated transmission spectrum, $y_{true}$ and $y_{pred}$ are the discrete data points generated by utilizing the ANNs and FDTD method, respectively. The best and worst possible values of the score are 1.0 and arbitrary negative, respectively.

It should be noted that the network architecture and the selection of the hyper-parameters determine the performance (prediction accuracy, convergence and calculation time) of ANNs [35]. It is generally true that a high computation cost is taken to train the deep neural networks due to the existence of a huge number of weights between the neurons in different layers [45]. In order to ensure good accuracy and reduce training time, the GA is applied in optimizing the network architecture and selecting the hyper-parameters (the algorithmic details of the GA are described in APPENDIX C). In the GA, the network architectures are full-connected, and four critical hyper-parameters (number of layers, neurons per layer, the solvers for weights and the activation functions for hidden layers) are regarded as the genetic genes. The score $1-J^2$ on the test sets is used as the fitness to evaluate each population's accuracy. As shown in Fig. 5(b), the scores are increased evolutionally and levelling out at high levels, which indicates the optimizations for ANNs are efficient. After optimizing the network architectures based on the GA, the suitable hyper-parameters for the THRC, FORC and FIRC systems are [8-200-400-300-300-300-50-200-200, 'relu', 'adam'], [12-400-200-300-400-100-200-200, 'tanh', 'adam'] and [15-300-400-400-200-400-200, 'relu', 'sgd'], respectively. Here, the input layers in the ANNs are the number of structure parameters, while the output layers match the discrete data points uniformly sampled from the transmission spectrum.

Due to the relatively simple network architecture, it takes a few minutes to train the ANNs by using the multi-layer perception regressor (MLPRegressor) in Scikit-learn library which is a famous machine learning toolbox for Python [48]. The other hyper-parameters, such as L2 penalty, batch_size, max_iter, and tolerance, are set to $10^{-5}$, 'auto', 1000, and $10^{-5}$ for all the PWCCSs. As shown in Fig. 5(b) and Fig. 5(c), for the THRC system, the score $1-J^2$ on the test sets is finally stabilized at 0.9862 and the training loss has sharp declines occasionally. It means that no matter training sets or test sets, the predicted transmission spectrums generated from the ANNs are very close to the simulation results calculated by FDTD method. To illustrate the effectiveness of the spectrum prediction based on the ANNs, arbitrary structure parameter is randomly selected from the test sets to make a comparative analysis between the ANNs predicted results and the FDTD simulation results. In Fig. 5(d), the red line relates to the FDTD simulated transmission spectrum corresponding to the structure parameters ($l_1$=466 nm, $l_2$=524 nm, $l_3$=589 nm, $w_1$=115 nm, $w_2$=93 nm, $w_3$=90 nm, $g_1$=280 nm, $g_2$=335 nm), while the blue dots represent that predicted by the ANNs for the same structure parameters. It can be observed that the double PIT effects predicted by the ANNs matches quite well with the FDTD simulated transmission spectrum even outside of the training sets. Obviously, the trained ANNs not only fit to the training data, but also learn some potential relationships between the structure parameters and the transmission spectrum for the THRC system. Similarly, the ANNs are also applied in spectrum prediction for the FORC and FIRC systems, and the comparison results are shown in Fig.5 (e) and Fig.5 (f), respectively. After many iterative rounds of model training, the scores on the test sets gradually raise to 0.9010 (FORC) and 0.9538 (FIRC), which indicates that the ANNs can effectively predict the transmission spectrums for the relatively complex PWCCSs. In Fig. 5(e) and Fig. 5(f), the ANNs-predicted transmission spectrums and the FDTD simulated transmission spectrums are broadly similar though the similarity for the steep optical characteristics (such as the FR) is imperfect. The reason for this imperfect is attributed to the insufficiency of the training data and the relatively simple network architectures. Actually, we can improve the precision of spectrum prediction by adding training data or designing complex network architecture. However, it is at the cost of training time and power, and the over-fitting problem is difficult to avoid [49, 50].

In addition, when the physical phenomena in the PWCCSs are theoretically analyzed, there are many theoretical parameters needed to be addressed by using data fitting method. It is more beneficial to automatically determine the theoretical parameters for specific electromagnetic response because the data fitting is an empirical and tedious process. We use ANNs to search the suitable parameters for the theoretical models in APPENDIX B, and it consists of the following steps: (i) 20000 training instances which includes the theoretical parameters and 200 discrete data points in theoretically calculated transmission spectrum are generated by utilizing MC method. It only takes a few seconds to generate the training sets because the computing process for theoretical models are not complex. (ii) In order to optimize the network architectures of the ANNs, we also use the GA to select the suitable hyper-parameters. In Fig.5 (g), it can be observed that the evolutionary scores are maintained at a higher level from the first generation because the theoretical models behind the physical phenomenon are really existed. (iii) We select three excellent ANNs whose scores on the test sets are greater than 99.60% to predict the fitting parameters for the FDTD simulated transmission spectrums, and the inset in Fig.5 (g) reveals the variation tendency of the loss in the model training. In Fig.2 (a), Fig.3 (a) and Fig.4 (a), the similarity between the theoretically calculated transmission spectrums and the FDTD transmission spectrums demonstrates the ANNs can predict the fitting parameters for the theoretical models.

For the PWCCSs shown in Fig. 1, the inverse design based on ANNs are also analyzed here. For this purpose, we should design arbitrary transmission spectrum within reasonable limits, and the ANNs could predict the structure parameters that would most closely produce the artificial transmission spectrum. Compared with the 'forward' ANNs which has applications in the spectrum prediction (from structure parameters to transmission spectrum), an 'inverse' network architecture which reproduce the structure parameters from the transmission spectrum is constructed specially. As shown in Fig.6 (a), the inputs and outputs of the 'inverse' network architecture are the discrete points uniformly sampled from the transmission spectrum and the structure parameters of the PWCCSs, respectively. Similarly, the 'inverse' ANNs are trained by using the 20000 training instances, and the network architectures are optimized by utilizing the GA. After a few iterative evolution steps, the suitable network architectures and hyper-parameters of the 'inverse' ANNs for the THRC, FORC and FIRC systems are [200-200-400-400-8, 'relu', 'sgd'], [200-200-400-300-100-12, 'relu', 'adam'] and [200-300-300-300-200-15, 'relu', 'adam'], respectively. Compared with the THRC system, the inverse design for the FORC and FIRC systems requires more sophisticated network architecture because there are more structure parameters should be predicted. The effectiveness of the inverse design for the THRC, FORC and

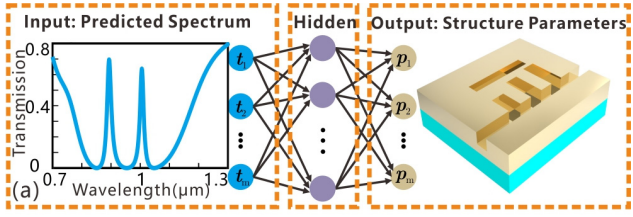

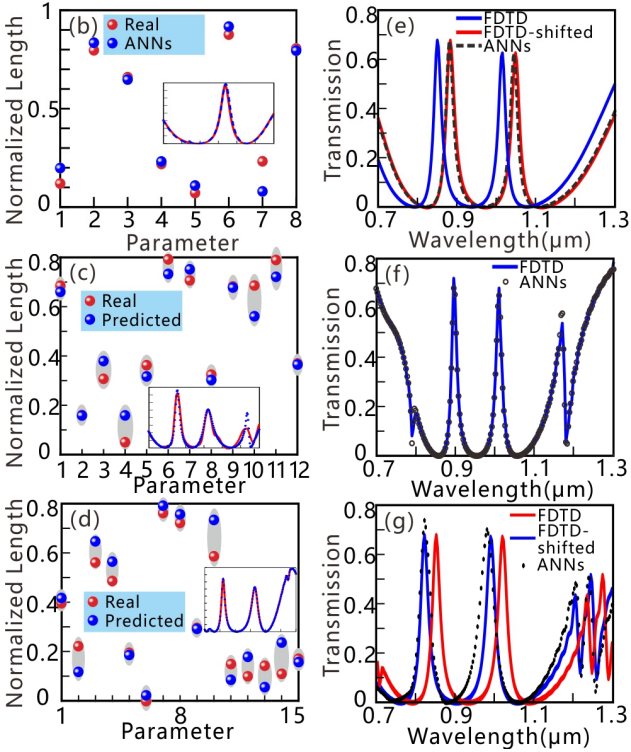

**Fig.6** (a) The diagram of the ANNs applied in the inverse design and performance optimization problems. The comparison results between the real structure parameters and ANNs-predicted structure parameters for the THRC (b), FORC (c) and FIRC (d) systems. The insets in (b)-(d) are the FDTD simulated transmission spectrums corresponding to the real structures (red solid line) and ANNs-predicted structure parameters (blue dash line). (e) The transmittance optimization for the FHRC system. (f) The bandwidth optimization for the FORC system. (g) The transmittance optimization for the FIRC system.

FIRC systems is quantitatively validated by calculating the score on the test sets. After a few iterative training steps, the score reaches to 0.912, 0.943 and 0.896 for the THRC, FORC and FIRC systems respectively. In order to provide a vivid visualization of the inverse design effect for the PWCCSs, the FDTD simulated transmission spectrums randomly selected from the test sets are inputted into the ANNs. The red circles in Fig. 6(b)-(d) show the real structure parameters, while the blue circles relate to the 'inverse' ANNs-predicted structure parameters. For the sake of convenience, the structure parameters are normalized to a range from 0 to 1. Interestingly, it can be observed that most of the predicted structure parameters agree with the real structure parameters accurately. To consider the influence of prediction error, the insets in Fig. 6(b)-(d) depict the FDTD simulated transmission spectrums corresponding to the real structure parameters (red lines) and predicted structure parameters (blue dots) for the THRC, FORC and FIRC systems, respectively. Compared with the FDTD results, it can be found that the structure parameters predicted by the ANNs can reproduce transmission spectrum with a high similarity. Obviously, it no doubt provides a new way to train ANNs for the inverse design of the PWCCSs.

Similar to the inverse design, the ANNs can be applied in optimizing for a specific property of the PWCCSs, such as transmittance, bandwidth, full width at half maxima (FWHM) and so on. In order to validate the performance optimization of the transmittance for an arbitrary wavelength and avoid generating the unreasonable results, the transmission spectrum randomly selected from the test sets is shifted manually for the THRC system. The blue solid line and red solid line in Fig. 6(e) are the FDTD simulated transmission spectrum and the red-shifted transmission spectrum, respectively. It can be observed that the transmittance at 900 nm increases from 0.05 to 0.68 by shifting the transmission spectrum. The red-shifted transmission spectrum is inputted into the 'inverse' ANNs, and the most probable structure parameters are predicted by the ANNs. The black dashed line in Fig. 6(e) represents the FDTD simulated result corresponding to the structure parameters predicted by the 'inverse' ANN ($l_1$=486 nm, $l_2$=550 nm, $l_3$=608 nm, $w_1$=89.7 nm, $w_2$=96 nm, $w_3$=94 nm, $g_1$=290 nm, $g_2$=341 nm). Obviously, the transmittance optimization for a given wavelength can be achieved by using ANNs due to the similarity between the ANNs-predicted transmission spectrum and the red-shifted transmission spectrum. Moreover, we try to optimize the bandwidth of the optical channel in the double PIT effects or FR based on the ANNs. For the FORC system, we expect to further reduce the bandwidth of the FR to achieve much steeper optical characteristics. For this purpose, the transmission spectrum is designed optimally (blue line in Fig. 6(f)), especially for the bandwidth of the FR (the bandwidth between the peak and dip of the FR is reduced from 12 nm to 8 nm). Then, the optimized transmission spectrum is inputted into the ANNs, and the predicted structure parameters are $l_1$=482 nm, $l_2$=539 nm, $l_3$=601 nm, $l_4$=903 nm, $w_1$=101 nm, $w_2$=100 nm, $w_3$=102 nm, $w_4$=101 nm, $g_1$=276 nm, $g_2$=331 nm, $g_3$=20 nm and $s_1$=1 nm. Here, the black dots in Fig. 6(f) represents the FDTD simulation results calculated for the predicted structure parameters. As shown in Fig. 6(f), the FDTD simulated results are close to the optimized transmission spectrum, which indicates the feasibility for bandwidth optimization by using the ANNs. Besides, the transmittance of the transmission spectrum for the FIRC system is also optimized. The red line in Fig. 6(g) is the original transmission spectrum randomly selected from the test sets for the FIRC system, and the blue line is the manually blue-shifted transmission spectrum. Here, the transmission spectrum in a given wavelength range (700 nm-1300 nm) can be shifted to achieve steeper optical characteristics or higher transmittance. When the blue-shifted transmission spectrum is inputted into the 'inverse' ANNs, the structure parameters ($l_1$=443 nm, $l_2$=526 nm, $l_3$=609 nm, $l_4$=896 nm, $l_5$=908 nm, $w_1$=104 nm, $w_2$=101 nm, $w_3$=109 nm, $w_4$=96 nm, $w_5$=111 nm, $g_1$=266 nm, $g_2$=327 nm, $g_3$=15 nm, $g_4$=60 nm, $s_1$=5.4 nm, $s_2$=7.4 nm) are predicted quickly. Apparently, the blue-shifted transmission spectrum agrees well with the FDTD simulated transmission spectrum (black dots line) calculated for the predicted structure parameters, which realizes the transmittance optimization for a specific wavelength in the FIRC system.

## 4. CONCLUSION

In this article, we propose a novel method using ANNs to achieve spectrum prediction, inverse design and performance optimization for the PWCCSs. The FR and PIT effects originated from the mode coupling in the PWCCSs are explained theoretically and taken as the example to verify the effectiveness of ANNs. The uncertain parameters in the theoretical models are fitted by using the ANNs effectively. In order to ensure good accuracy and reduce the training time, we use the GA to design the network architectures and select the suitable hyper-parameters for ANNs. It is important to note that the transmission spectrum predicted by ANNs are approximate to the FDTD simulated results with high precision. More importantly, the ANNs have been successfully employed in solving the inverse design and performance optimization problems for the PWCCSs. Obviously, we construct a novel model-driven analysis method for the PWCCSs which are expected to have significant applications in the design, analysis and optimization for optical devices.

## APPENDIX A. STRUCTURE PARAMETERS AND SIMULATION SETTINGS

In all PWCCSs, silver (Ag) is selected as the background metals which support the propagation of SPPs (the relative permittivity is described by using Drude model with ($\varepsilon_\infty$, $\omega_p$, $\gamma_p$) = (3.7, 9.1 eV, 0.018 eV)) [6-9]. The relative permittivity of air $\varepsilon_a$ is 1. We have calculated the transmission loss factor of the SPPs mode in the MDM waveguide, and the results are shown in Fig.2 (b). It should be noted that the maximum loss factor in our operating wavelength region is only 105.4 dB/mm at $\lambda$=700 nm, resulting a small loss (0.93dB) for SPPs propagating from the cavity1 to cavity 3 in the THRC system. In addition, as shown in Fig. 2(a), the loss of propagated SPPs mode would decrease the transmission, but it didn't influence the position of the resonance wavelength of three comb cavities. As a result, Ag is a suitable material for propagation of SPPs in our proposed PWCCSs. The width of the bus waveguide is fixed as $w$=100 nm. The structure parameters of the cavities in the PWCCSs are stochastic variable within certain range, and we only introduce the initial values: the lengths and widths of all comb cavities 1, 2 and 3 are kept as $l_1$=480 nm, $l_2$=520 nm, $l_3$=560 nm and $w_1$=$w_2$=$w_3$=100 nm, respectively. The waveguide coupling distances between the comb cavities 1, 2 and 3 are $g_1$=275 nm and $g_2$=330 nm. In Fig. 1(b), the coupling distance between the rectangular cavity 4 and comb cavity 3 is $g_3$=20 nm, and the length and width of the rectangular cavity 4 are set to be $l_4$=$w_1$+$w_2$+$w_3$+$g_1$+$g_2$=885 nm and $w_4$=100 nm, respectively. In Fig. 1(c), the structure parameters of the rectangular cavity 5 are similar to those of the cavity 4, and the coupling distance between the cavities 4 and 5 is $g_4$= 20 nm.

The characteristic spectral responses of all the PWCCSs in this article are calculated by using the FDTD method (adopting the commercial software Lumerical FDTD Solutions). In the actual experiment, the incident laser source can excite SPPs wave via a metal grating located at the front of bus waveguide [14-16]. Since the SPPs waves are intensely confined to the metal–air interface, the PWCCSs are simplified to 2D model for numerical simulation [6-9]. The incident source is the solved eigen-mode by using FDTD Solutions, and the distance between the source and monitor is 1100 nm. The perfectly matched layers are used for the simulation boundary conditions, and the number of the perfectly matched layer is 8. In order to ensure the accuracy and algorithm convergence, the mesh settings in cavities are 20 grids in width direction and 80 grids in length direction, while those in waveguide are 30 grids in width direction and 200 grids in length direction. The non-uniform meshes with an 8-level mesh accuracy are adopted to represent the other simulation regions.

## APPENDIX B. THEORETICAL ANALYSIS OF THE FR AND PIT EFFECTS

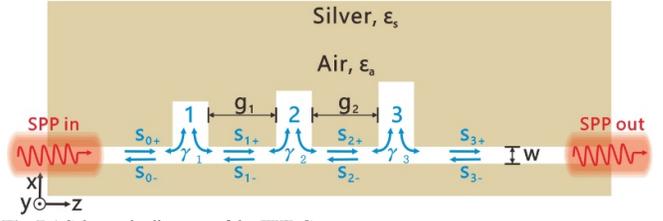

**Fig. B1** Schematic diagram of the THRC system.

To obtain a qualitative understanding of the physical phenomena, we theoretically analysis the FR and PIT effects originated from the PWCCSs by combining the CMT and TMM [17]. As shown in Fig. B1, the coupling coefficients between the plasmonic waveguide and three cavities are $\gamma_1$, $\gamma_2$ and $\gamma_3$, respectively. And the amplitudes of the input and output waves into the cavities are donated by $s_{m+}$, $s_{m-}$ ($m$=0, 1, 2, 3). Since cavities 1, 2 and 3 don't directly couple with each other, we take a simple case into consideration where only a single cavity connects to the bus waveguide. The temporal change of the normalized mode amplitude of the cavity $a_j$ is described by

$$\frac{da_m}{dt} = (j\omega_m - \gamma_m)a_m + j\sqrt{\gamma_m}\left(s_{(m-1)+} + s_{m-}\right) \quad (B1)$$

where $\omega_m$ ($m$=1, 2, 3) is the resonant frequency of the cavities 1, 2 and 3, and time dependence is assumed exp($j\omega t$). Due to the energy conversation and the time reversal symmetry, we can derive the relationships between the amplitudes of the input-output waves in the waveguide and the resonant modes in cavities 1, 2 as follows

$$s_{(m-1)-} = s_{m-} + j\sqrt{\gamma_m}a_m \quad (B2)$$

$$s_{m+} = s_{(m-1)+} + j\sqrt{\gamma_m}a_m \quad (B3)$$

Using the Eqs. (B1)-(B3), it can be derived that

$$s_{(m-1)-} = -\frac{\gamma_m}{p_m - \gamma_m}s_{m-} + \frac{p_m - 2\gamma_m}{p_m - \gamma_m}s_{m+} \quad (B4)$$

$$s_{(m-1)+} = \frac{p_m}{p_m - \gamma_m}s_{m-} + \frac{\gamma_m}{p_m - \gamma_m}s_{m+} \quad (B5)$$

where $p_m$=j($\omega$-$\omega_m$)+$\gamma_m$, Eqs. (B4) and (B5) can be written in the transfer matrix form

$$\begin{pmatrix} s_{(m-1)-} \\ s_{(m-1)+} \end{pmatrix} = M_m \begin{pmatrix} s_{m-} \\ s_{m+} \end{pmatrix} \quad (B6)$$

where

$$M_m = \frac{1}{p_m - \gamma_m}\begin{pmatrix} p_m - 2\gamma_m & -\gamma_m \\ \gamma_m & p_m \end{pmatrix} \quad (B7)$$

For the THRC system shown in Fig. B1, the relationship between input and output waves can be expressed as

$$\begin{pmatrix} s_{0-} \\ s_{0+} \end{pmatrix} = M_1 \begin{pmatrix} e^{j\varphi_1} & 0 \\ 0 & e^{-j\varphi_1} \end{pmatrix} M_2 \begin{pmatrix} e^{j\varphi_2} & 0 \\ 0 & e^{-j\varphi_2} \end{pmatrix} M_3 \begin{pmatrix} s_{3-} \\ s_{3+} \end{pmatrix}$$
$$(B8)$$

where $\varphi_n$=$k_0 n_{MDM} g_n$ ($n$=1, 2) is the phase shift induced by the SPPs propagating between two adjacent cavities, and $n_{MDM}$ is the effective index of waveguide which is solved by the dispersion equation [9]

$$\tanh\left(\frac{w\pi\sqrt{n_{MDM}^2-\varepsilon_a}}{\lambda}\right)=-\frac{\varepsilon_a\sqrt{n_{MDM}^2-\varepsilon_m}}{\varepsilon_m\sqrt{n_{MDM}^2-\varepsilon_a}} \quad (B9)$$

The transfer matrix of $M_{all}^{THRC}$ for the THRC system is expressed as

$$M_{all}^{THRC}=M_1\begin{pmatrix}e^{j\varphi_1}&0\\0&e^{-j\varphi_1}\end{pmatrix}M_2\begin{pmatrix}e^{j\varphi_2}&0\\0&e^{-j\varphi_2}\end{pmatrix}M_3 \quad (B10)$$

Since we only launch SPPs mode into the system from the input port, i.e. $s_{3-}=0$, the transmission of the THRC system is determined by $|1/M_{all,22}^{THRC}|^2$. Using Eqs. (B7) and (B10), we can obtain the transmission of the THRC system

$$T^{THRC}=\left[\frac{(p_1-\gamma_1)(p_2-\gamma_2)(p_3-\gamma_3)}{\gamma_1\gamma_3(2\gamma_2-p_2)e^{j(\varphi_1+\varphi_2)}-p_1\gamma_2\gamma_3e^{j(\varphi_2-\varphi_1)}-\gamma_1\gamma_2p_3e^{j(\varphi_1-\varphi_2)}+p_1p_2p_3e^{-j(\varphi_1+\varphi_2)}}\right]^2 \quad (B11)$$

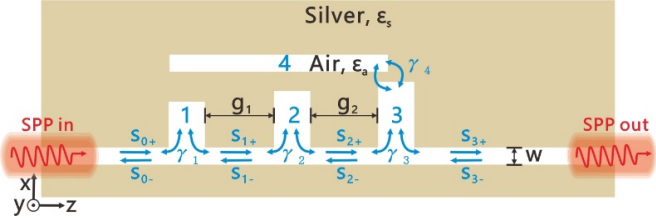

**Fig. B2** Schematic of the FORC system.

As shown in Fig. 3, the FR is induced by the coupling between the cavity 3 and cavity 4, and the coupling between the cavities 1, 2 and cavity 4 is very weak, thus we can only alter the transfer matrix $M_3$ in Eq. (B10) to obtain the transmission spectrum of the FORC system. We assume the resonant frequency of the cavity 4 is $\omega_4$. The time evolution of the amplitudes of the cavities 3 and 4 in steady state can be described as

$$\frac{da_3}{dt}=(j\omega_3-\gamma_3-\gamma_4)a_3 \quad (B12)$$
$$+j\sqrt{\gamma_3}(s_{2+}+s_{3-})+j\sqrt{\gamma_4}a_4$$

$$\frac{da_4}{dt}=(j\omega_4-\gamma_4)a_4+j\sqrt{\gamma_4}a_3 \quad (B13)$$

Using Eqs. (B2), (B3), (B12) and (B13), the transfer matrix $M_3$ for the FORC system can be derived as

$$M_3^{FORC}=\frac{1}{p_3^{FORC}-\gamma_3}\begin{pmatrix}p_3^{FORC}-2\gamma_3&-\gamma_3\\\gamma_3&p_3^{FORC}\end{pmatrix} \quad (B14)$$

where $p_3^{FORC}=j(\omega-\omega_3)+\gamma_3+\gamma_4+\gamma_4/[j(\omega-\omega_4)+\gamma_4]$. Thus, employing Eqs. (B7), (B10) and (B14), the transmission of the FORC system can be expressed as

$$T^{FORC}=\left[\frac{(p_1-\gamma_1)(p_2-\gamma_2)(p_3^{FORC}-\gamma_3)}{\gamma_1\gamma_3(2\gamma_2-p_2)e^{j(\varphi_1+\varphi_2)}-p_1\gamma_2\gamma_3e^{j(\varphi_2-\varphi_1)}-\gamma_1\gamma_2p_3^{FORC}e^{j(\varphi_1-\varphi_2)}+p_1p_2p_3^{FORC}e^{-j(\varphi_1+\varphi_2)}}\right]^2 \quad (B15)$$

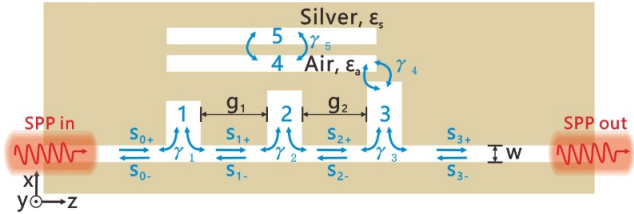

**Fig. B3** Schematic of the FIRC system.

Similarly, for the FIRC system illustrated in Fig. B3, the time evolution of the mode amplitudes of the cavities 3, 4 and 5 in steady state can be described as

$$\frac{da_3}{dt}=(j\omega_3-\gamma_3-\gamma_4)a_3 \quad (B16)$$
$$+j\sqrt{\gamma_3}(s_{2+}+s_{3-})+j\sqrt{\gamma_4}a_4$$

$$\frac{da_4}{dt}=(j\omega_4-\gamma_4-\gamma_5)a_4+j\sqrt{\gamma_4}a_3+j\sqrt{\gamma_5}a_5 \quad (B17)$$

$$\frac{da_5}{dt}=(j\omega_5-\gamma_5)a_5+j\sqrt{\gamma_5}a_4 \quad (B18)$$

Using Eqs. (B2), (B3) and (B16)-(B18), the transfer matrix $M_3$ for the FIRC system can be derived as

$$M_3^{FIRC}=\frac{1}{p_3^{FIRC}-\gamma_3}\begin{pmatrix}p_3^{FIRC}-2\gamma_3&-\gamma_3\\\gamma_3&p_3^{FIRC}\end{pmatrix} \quad (B19)$$

where

$$p_3^{FIRC}=j(\omega-\omega_3)+\gamma_3+\gamma_4$$
$$+\cfrac{\gamma_4}{j(\omega-\omega_4)+\gamma_4+\gamma_5+\cfrac{\gamma_5}{j(\omega-\omega_5)+\gamma_5}}$$

(B20)

Thus, employing Eqs. (B7), (B9), (B10) and (B20), the transmission of the FIRC system can be obtained

$$T^{FIRC}=\left[\frac{(p_1-\gamma_1)(p_2-\gamma_2)(p_3^{FIRC}-\gamma_3)}{\gamma_1\gamma_3(2\gamma_2-p_2)e^{j(\varphi_1+\varphi_2)}-p_1\gamma_2\gamma_3e^{j(\varphi_2-\varphi_1)}-\gamma_1\gamma_2p_3^{FIRC}e^{j(\varphi_1-\varphi_2)}+p_1p_2p_3^{FIRC}e^{-j(\varphi_1+\varphi_2)}}\right]^2 \quad (B21)$$

# APPENDIX C. DESIGN OF THE NEURAL NETWORK ARCHITECTURE

In this article, the genetic algorithm (GA) is used in designing the network architecture and selecting the hyper-parameters of ANNs. The GA consists of the following steps: (i) randomly generating $N$=20 network architectures to create initial populations as the first generation. Here, four critical hyper-parameters including number of layers (3, 4, 5, 6, 7, 8), neurons per layer (10, 50, 100, 200, 300, 400), the solvers for weight optimization (sgd, adam) and the activation functions for the hidden layer (relu, tanh) are regarded as the genetic genes. The network architectures are constructed by selecting random values for the above-mentioned hyper-parameters. (ii) Evaluating each population's fitness. The test sets' score which measures the distance between the results predicted by the ANNs and the ground truth is regarded as the fitness function. It takes some time because we have to train the weights for each network and see how well it performs on the test sets. (iii) If the generation of networks evolves for 10 times or the fitness does not increase for more than 3 generations, then the optimization process is stopped; otherwise, proceed to Step (iv). (iv) New population consisted of new network architectures is reproduced and updated by selecting, crossing and mutating the genetic genes based on each population's fitness. In the process of selection, the networks architectures in current population are sorted by fitness. We keep some percentage of the top networks (25%, 5 networks) to become part of the next generation and to reproduce children. In addition, we also randomly keep 3 more loser networks and mutate a few of them to avoid falling into local optimum. In the process of crossover, two children network architectures replace their parent network architectures by combining the hyper-parameters randomly from their parents. For instance, one network architecture might have the same number of layers as its father and the rest of its parameters from its mother. In order to add randomness, each population in the new generation has 5% probability of mutation (a hyper-parameter is randomly changed to other value in the choice space). Then, an algorithmic loop is constructed by evaluating the new generation in Step (ii), judgment in Step (iii) and reproducing new population in Step (iv) [24].

**Funding.** National Natural Science Foundation of China (Grant No. 61625104, No. 61431003); China Postdoctoral Science Foundation (Grant No. 2017M610826, No. 2018T110074); National Key Research and Development program (Grant No.2016YFA0301300); The Beijing Municipal Science & Technology Commission (Grant No. Z181100008918011).